\documentstyle[aps,draft,floats]{revtex}
\tightenlines
\input{epsf}
\begin{document}
\title{Steady-State Cracks in Viscoelastic Lattice Models II}
\author{David A. Kessler\cite{barilan}}
\address{Dept. of Mathematics, Lawrence Berkeley National Laboratory,
1 Cyclotron Road, Berkeley, CA 94720}
\maketitle
\begin{abstract}
We present the analytic solution of the Mode III steady-state crack
in a square lattice with piecewise linear springs and Kelvin viscosity.
We show how the results simplify in the limit of large width.  We relate
our results to a model where the continuum limit is taken only along the 
crack direction.  We present results for small velocity, and for large
viscosity, and discuss the structure of the critical bifurcation for small
velocity. We compute the size of the process zone wherein standard
continuum elasticity theory breaks down.

\end{abstract}

\section{Introduction}
The problem of the dynamics of cracks has received renewed interest
recently,\cite{review} motivated in large part by new sets of experiments.\cite{texas,fineberg}
These experiments have
called into question some of the predictions of the traditional,
continuum mechanics approach to fracture dynamics. The most striking
experimental finding is that cracks exhibit a branching instability
long before they reach the predicted limiting speed of advance.  This
instability causes increased dissipation and sets an effective limit
on the speed of crack propagation.  There are hints of such an instability
in the continuum approach,\cite{yoffe} but a systematic treatment remains elusive.\cite{langer-recent}

One avenue of exploration that has proven fruitful is the lattice
models of fracture pioneered by Slepyan\cite{slepyan,slepyan2} and 
further developed by
Marder and collaborators.\cite{marder_gross,marder-liu}  
These models, especially in the
extreme brittle limit,  are simple enough to allow
comprehensive study, both analytically and by numerical simulation.
The lattice models exhibit some novel effects, not seen in the
continuum description.  Foremost is the existence of arrested cracks.
The lattice models also show instabilities at large velocities
that may be relevant to the experimentally seen branching instabilities.
Thus, it is useful to understand the lattice models in as much detail as
possible.

In a previous paper,\cite{kl_crack1} we embarked on a study of the effect of dissipation,
in the form of a Kelvin viscosity,\cite{langer} on the behavior of steady-state cracks.
We solved numerically for the dependence of velocity as a function of the
driving displacement $\Delta$.  We found that dissipation acts to lower the
velocity and significantly reduces the size of the lattice-induced
small velocity unstable regime where the velocity is a decreasing function
of the driving. We also showed that in the presence of dissipation,
the stable regime is well approximated by a novel $x$-continuum model,
wherein the lattice structure perpendicular to the crack is retained
but along the crack is replaced by a naive continuum limit.  We also
showed that if the transverse dimension $N$ is large, then
at distances of order $N$ the
elastic fields are given by the results of standard continuum fracture
theory.  On small scales, however, there is a boundary layer where
the discreteness of the lattice in the transverse direction is important.
This boundary layer structure is all important in determining the velocity
versus driving relation.  However, as our $x$-continuum model demonstrated,
the discreteness in the direction of the crack is less crucial, and primarily
affects the small velocity regime.

In this current paper, we study the large-$N$ limit of the theory.  We
do this first for our $x$-continuum model, where the structure of the
theory is simpler.  We then extend this to the full lattice model.
In both cases, we present a formal Wiener-Hopf solution of the model
for arbitrary $N$, and then take the large-$N$ limit.  This is in
contrast to the work of Slepyan, who, for the case of infinitesimal
dissipation, solves the infinite-$N$ limit directly.  The principal
advantage of our method is that it allows a discussion of case
of large, but finite, $N$.  It also allows a comparison between
the small-scale and the large-scale structure, whereas Slepyan's
method only produces a solution for small to intermediate scales.  Thus
Slepyan must rely on an implicit matching to large scales via the
stress-intensity factor, as opposed to the explicit matching contained
in our solution.  The Slepyan method, nevertheless, by avoiding the 
neccesity of solving the finite-$N$ problem, is more easily applied to other
cases, such as the mode-I problem, where the finite-$N$ solution is not
so easily obtained.

The plan of the paper is as follows.  In Section II, we describe the
lattice model and the simpler $x$-continuum version. In Section III, we
lay out our major results.  The details of the calculation are contained
in the following sections, first for the $x$-continuum problem in Section IV,
and then for the lattice problem in Section V.  The small velocity limit
is studied in Section VI and the large viscosity limit in Section VII.
We conclude with some comments in Section VIII.

\section{Description of the Models}

The lattice model we study is identical to that described in our
earlier work, \cite{kl_crack1}.  We have a square lattice of mass points
undergoing (scalar) displacement out of the plane. The lattice extends
 infinitely long in the $x$-direction, with $N+1$ rows in the 
$y$-direction.  The
lattice points are connected by linear ``springs'', with
spring constant 1, to their nearest
neighbors.  The top row is displaced a fixed amount $\Delta$.  The
bottom row is connected to a fixed line, with piece-wise linear springs.
These springs, with spring constant $k$, ``crack'' irreversibly if they 
are stretched an amount $\epsilon$.  When $k=2$, this model is equivalent to a
system of $2N+2$ rows, loaded by $\pm\Delta$ from top
and bottom, with a symmetric crack running down the middle, with
extension at cracking of the springs that bridge the middle being
$2\epsilon$. All the
(uncracked) springs have a viscous damping $\eta$.  The equation of 
motion for the system is then
\begin{equation}
{\ddot u}_{i,j} = \left(1 + \eta \frac{d}{dt}\right) (u_{i+1,j} + u_{i-1,j} +
u_{i,j+1} + u_{i,j-1} - 4u_{i,j})
\end{equation}
for $j \ne 1$ with $u_{i,N+1}\equiv\Delta$, and
\begin{equation}
{\ddot u}_{i,1} = \left(1 + \eta \frac{d}{dt}\right) (u_{i+1,j} + u_{i-1,j} + u_{i,2}
- 3u_{i,j}) - k \theta(\epsilon - u_{i,1} ) 
\left(1+\eta\frac{d}{dt}\right)u_{i,1} \ .
\end{equation}
Note that in these units, the elastic wave speed is unity, so all
velocities are dimensionless, expressed as fractions of the wave speed.

We are interested in steady-state cracks, described by the Slepyan
traveling wave ansatz,
\begin{equation}
u_{i,j}(t) = u_j(t-i/v)
\end{equation}
which implies that every mass point in a given row undergoes the
same time history, translated in time.  We choose the origin at
time such that $u_1(0)=\epsilon$ so that it represents the moment
of cracking of the spring attached to the bottom row mass point.
The equation of motion is best expressed in terms of the $N\times N$ coupling
matrix
\begin{equation} 
\label{m_define}
{\cal M}_N(m) = \left[
\begin{array}{ccccccc}
 -(m+1)&   1 &     &         &    &    &\\
   1   & -2 &   1  &         &    &    &\\
       &   1 & -2  &   1     &    &    &   \\
       &     &     &  \ddots &    &    &   \\
       &     &     &         &  1 & -2 & 1 \\
       &     &     &         &    &  1 & -2
\end{array}
\right ]
\end{equation}
The steady-state equation then reads
\begin{eqnarray}
\ddot u_j(t) &-& \theta(t)\left(1+\eta\frac{d}{dt}\right){\cal M}_{j,j'}(0)u_{j'}(t)
- \theta(-t)\left(1+\eta\frac{d}{dt}\right){\cal M}_{j,j'}(k)u_{j'}(t) \nonumber\\
 &-& \left(1+\eta\frac{d}{dt}\right)(u_j(t+1/v) -2u_j(t)+ u_j(t-1/v))=0
\end{eqnarray}

We will also consider in this paper an $x$-continuum version of this
model, where we replace the nonlocal in time coupling 
along the crack with its continuum analog
\begin{eqnarray}
\label{cont_model}
\ddot u_j(t) &-& \theta(t)\left(1+\eta\frac{d}{dt}\right){\cal M}_{j,j'}(0)u_{j'}(t)
- \theta(-t)\left(1+\eta\frac{d}{dt}\right){\cal M}_{j,j'}(k)u_{j'}(t) \nonumber\\
 &-& \left(1+\eta\frac{d}{dt}\right){\frac{1}{v^2}\ddot u_j(t)} = 0
\end{eqnarray}

\section{Survey of Results}
In this section, we survey the major results derived in the bulk of the
paper.  As the derivations are exceedingly technical, it is useful
to present the results first by themselves so that they may
be appreciated without getting lost in a welter of technical complications.

We begin by completing the Wiener-Hopf (WH) solution of the continuous $x$,
discrete $y$ model, as the results are simpler and are a useful
basis for assimilating the more complicated results of the full lattice
model.  The key aspect of the solution is the calculation
of $\Delta$ as a function of the crack velocity $v$ (in units where the
wave speed is unity).  We find
\begin{equation}
\frac{\Delta}{\Delta_G} = \sqrt{kN+1}\prod_m \frac{q_{1,m}(1+\eta v Q_{1,m})}{Q_{1,m}(1+\eta v q_{1,m})}
\end{equation}
which expresses $\Delta$ (normalized to the Griffith value
\begin{equation}
\Delta_G=\epsilon\sqrt{2N+1}
\end{equation}
at which the uncracked state becomes metastable) 
in terms of the wave vectors corresponding to the
various normal modes of the problem.  If we label the normal mode eigenvalues
of the $y$-coupling matrix on the uncracked side ${\cal M}(k)$ by
$\Lambda_m$, then $Q_{1,m}$ is the unique positive root of the dispersion
relation
\begin{equation}
\eta v Q^3 + (1-v^2)Q^2 + (1+\eta v Q)\Lambda_m = 0  \ .
\end{equation}
Similarly, $q_{1,m}$ is the unique positive root of the dispersion relation
using the normal mode eigenvalue $\lambda_m$ of the cracked side 
${\cal M}(0)$. 

This formula is fairly complicated, but simplifies tremendously for the
case of symmetric cracks ($k=2$) in the macroscopic limit $N>>1$. Then,
the product above can be performed analytically, with the simple result
\begin{equation}
\frac{\Delta}{\Delta_G} = (1-v^2)^{-1/4}
\sqrt{\frac{2(1+\eta v Q_1(1))}{Q_1(1)}}
\end{equation}
where $Q_1(1)$ is the mode associated with the highest frequency $y$-mode,
with $\Lambda=-4$.  For typical $\eta$'s of order 1, $Q_1(1)$ does not
vary much from its zero velocity value of 2.  
The resulting curve $\Delta(v)/\Delta_G$ starts
linearly at $v=0$ from 1 with slope $\eta$ and diverges at $v=1$.  Thus,
in the infinite $N$ limit, the velocity never exceeds the wave-speed.  At
any finite $N$, however, the velocity crosses the wave speed at a $\Delta$
of order $N^{1/6}\Delta_G$. Since the divergence with $N$ is so weak,  crossing the wave-speed barrier 
may not be as difficult as one would naively think.  This is especially
true for small dissipation, where the critical $\Delta$ scales as
$(\eta N)^{1/6}$.  This appears a more likely mechanism for explaining
the experimental observation of supersonic cracks than the time-dependent 
forcing hypothesis of Slepyan.\cite{slep-fast}

The basic structure is unchanged when we go over to the full lattice
model.  The essential difference is that the lattice dispersion relation
is nonpolynomial and has an infinite number of positive (real-part) solutions
for each eigenmode $m$.  The $\Delta - v$ relationship is
\begin{equation}
\label{final_cont_0}
\frac{\Delta}{\Delta_G} = \sqrt{kN+1}\prod_{n,m} \frac{q_{1,n,m}(1+\eta v Q_{1,n,m})}{Q_{1,n,m}(1+\eta v q_{1,n,m})}
\end{equation}
where now the product extends 
over all positive real-part roots $Q_{1,n,m}$ of the lattice 
dispersion relation
\begin{equation}
0 = (1+\eta v Q) (4\sinh^2(Q/2) + \Lambda_m) - v^2 Q^2
\end{equation}
for each $\Lambda_m$ ($\lambda_m$ in the case of $q_{1,n,m}$).  For a given
$m$, there is one real positive root, $Q_{1,0,m}$ ($q_{1,0,m}$), and an infinite series of complex-conjugate pairs of complex roots, ordered by increasing
imaginary part.  For large $n$, the imaginary part increases by roughly $2\pi$
for each successive root. 

Again, for symmetric cracks we can evaluate analytically the macroscopic
(large $N$) limit.  We obtain
\begin{equation}
\label{final_lat_0}
\frac{\Delta}{\Delta_G} = 
(1-v^2)^{-1/4}\sqrt{\frac{2(1+\eta v q_{\infty,0})}{q_{\infty,0}}}\left[\prod_
{n\ne 0}\frac{q_{0,n}(1+\eta v q_{\infty,n})}{q_{\infty,n}(1+\eta v q_{0,n})}\right]^{1/2}
\end{equation}
where $q_{\infty,n}$ is the root corresponding to the highest frequency
$\Lambda=-4$ eigenmode and plays the role of $Q_{1}(1)$ of the previous
$x$-continuum result.  The $q_{0,n}$ are the roots corresponding to the
$\Lambda=0$ eigenmode.  These do not have a counterpart in the $x$-continuum
calculation as the $n=0$ real solution vanishes, and only the lattice-induced
$n\ne 0$ modes enter.

\begin{figure}
\centerline{\epsfxsize=3.25in \epsffile{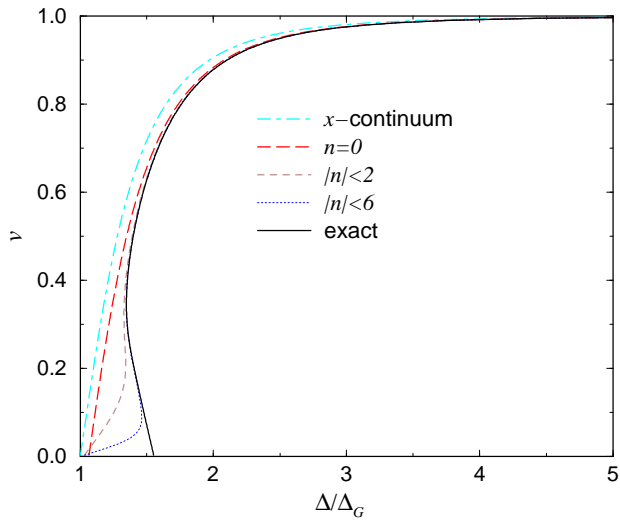}}
\caption{$v$ vs. $\Delta/\Delta_G$ in the $x$-continuum approximation,
Eq.\ (\ref{final_cont_0}), and in the exact lattice model, together with
the lattice result truncated after the $n=0$ term, $|n|=1$ and $|n|=5$ terms.}
\label{v_vs_delta_eta=.5_partial}
\end{figure}

As indicated by the way we expressed this result, we can consider it
as essentially the $x$-continuum result, Eq.\ (\ref{final_cont_0}), with
the real lattice $q_{\infty,0}$ replacing $Q_{1}(1)$, modified by a
multiplicative correction factor involving the complex lattice modes.
To understand the usefulness of this way of thinking, as well as its
limitations, we present in Fig. \ref{v_vs_delta_eta=.5_partial}, for $\eta=.5$, the exact numerically computed 
relationship Eq.\  
(\ref{final_lat_0}), along with the $x$-continuum result Eq.\  
(\ref{final_cont_0}).  In addition, we plot the lattice result, truncated
after its $n=0$, $|n|=1$ and $|n|=5$ terms. We see that at larger velocities
all these results are close, indicating that the lattice-induced shift
in $q$ as well as the additional lattice modes play little role at these
velocities.  At smaller velocities, the various approximations
differ significantly from each other and from the exact curve.  We see, in
fact, that as $v$ approaches 0, more and more terms must be included in 
the product to achieve an accurate result.  The calculation of the limiting
behavior at small velocities requires summing all the terms.  
The result of the calculation
is that for all $\eta$, as $v \to 0$, $\Delta$ approaches 
$\Delta|_{0^+}=\sqrt{1+\sqrt{2}}\Delta_G$, 
the maximal $\Delta$ for which an arrested crack exists.  This generalizes
the result of Slepyan for infinitesimal dissipation. As $v$ increases,
$\Delta$ decreases linearly with the $\eta$--independent slope, 
$-\Delta|_{0^+}/2$ so that the bifurcation from the arrested is
subcritical and universal.

\global\firstfigfalse

\begin{figure}
\centerline{\epsfxsize=3.25in \epsffile{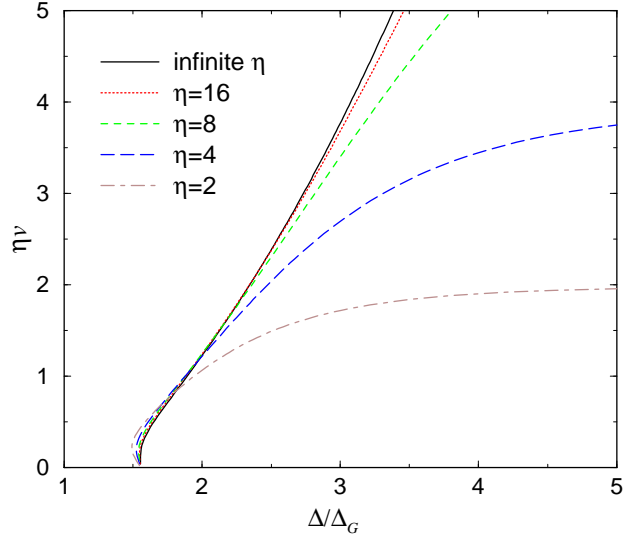}}
\caption{$\eta v$ vs. $\Delta/\Delta_G$ for $\eta=2, 4, 8, 16$
along with the asymptotic result for large $\eta$, Eq.\ (\ref{large_eta_0}).}
\label{large_eta_fig}
\end{figure}

More progress can be made in the large $\eta$ limit.  Here, at fixed
$\Delta$, the velocity goes to zero as $\eta$ increases, so that the
ratio $\phi\equiv\eta v$ is fixed.  In this limit, we can calculate the infinite
product and find
\begin{equation}
\label{large_eta_0}
\frac{\Delta}{\Delta_G} = \left[\coth(\frac{1}{2\phi}) + \sqrt{2}\right]^{1/2}
\end{equation}
This infinite-$\eta$ result, together with the exact result for various
$\eta$'s, is presented in Fig. \ref{large_eta_fig}.  

We see that this calculation does not reproduce the subcritical bifurcation
from the arrested crack at small velocities, which is a higher order effect.
We can evaluate this $1/\eta$ correction near the bifurcation at small $\phi$,
and find
\begin{equation}
\Delta \sim \Delta|_{0^+} (1 + \frac{\Delta_G^2}{\Delta|_{0^+}^2}e^{-1/\phi})(1 - \frac{\phi}{2\eta})
\end{equation}
which reproduces the small $v$ behavior described above and shows that
the $\eta$ dependent corrections are in fact exponentially small in $v$.
The resulting $\Delta - v$ curve starts at $\Delta|_{0^+}$ at $v=0$, heads
back linearly for a short distance of order $1/\eta (\ln\eta)^2$
and then sharply veers forward.

A last result worth noting is that whereas the Kelvin viscosity model
analyzed herein has a nice macroscopic limit when expressed in terms of
$\Delta_G$, the model with Stokes viscosity, where the dissipation in
put in the masses and not in the bonds, does not have such a limit.  There
an $O(1)$ Stokes viscosity at the microscopic level changes the continuum
elastic fields and requires an ever-increasing $\Delta/\Delta_G$ as the
sample is made wider.

\section{The \lowercase{$x$}--Continuum Model}
We begin our analysis with the solution of the $x$--continuum model, Eq.\ (\ref{cont_model}),
introduced in Kessler, et. al.~\cite{kl_crack1}. It is important to
remember that is this model, the lattice structure in the $y$-direction
is left unchanged.
The solution of the
lattice model is similar in structure to that of the $x$--continuum
model, but the latter is a
simpler context in which to develop the necessary techniques.  Furthermore,
the $x$--continuum model is an interesting approximation in its own
right, which captures a significant amount of the structure of the full
lattice problem.

In Kessler, et. al.~\cite{kl_crack1}, a Wiener--Hopf analysis of the
problem was initiated.  In this analysis, the key technique is to
decompose all the terms in the steady-state equation of motion
into terms analytic in the upper- and lower-half planes respectively.
However, the analysis was not carried to
completion, due to the presence of one term whose decomposition
was not evident.  Here we use a trick to accomplish the decomposition
of this last remaining term, and thereby complete the solution of the
problem.  We choose not to reproduce the lengthy preliminary
stages of this calculation, for which the interested reader
is referred to \cite{kl_crack1}.  We do however reiterate
the definition of the relevant notations introduced there,
so that the current exposition is minimally self-contained.

The problematic term, from \cite{kl_crack1} Eq.\ (42), is
\begin{equation}
ik\frac{\prod_l(K-i\chi_{1,l})(K+i\chi_{2,l})(K+i\chi_{3,l})}
{\prod_m(K-iq_{1,m})(K+iQ_{2,m})(K+iQ_{3,m})}
\end{equation}
where the $\chi$, $q$, and $Q$ are the roots of a certain family of
cubic polynomials.  In detail, let $\ell_l$, ($l=1,\ldots,N-1$) be
the eigenvalues of the $(N-1)\times (N-1)$ coupling matrix 
${\cal M}_{N-1}(1)$ defined in Eq.\ (\ref{m_define}) above.  Define
the polynomial specifying the dispersion relation, $P(\lambda,Q)$, by
\begin{equation}
P(\lambda,Q)=\eta v Q^3 + (1-v^2)Q^2 + (1+\eta vQ)\lambda
\end{equation}
Then, $P(\ell_m,Q)$ has, for each $m$, three roots, one positive which we
denote $\chi_{1,m}$, and two with negative real parts, which we
denote by $-\chi_{2,m}$, $-\chi_{3,m}$, so that all the $\chi$'s have
positive real parts.  Similarly, denote the eigenvalues of the
$N\times N$ matrix ${\cal M}_N(k)$ by $\Lambda_m$, $m=1,\ldots,N$.
Then $Q_{1,m}$, $-Q_{2,m}$, and $-Q_{3,m}$ are the roots of $P(\Lambda_m,Q)$.
Likewise, denote the eigenvalues of ${\cal M}(0)$ by $\lambda_m$.  Then
$q_{1,m}$, $-q_{2,m}$, and $-q_{3,m}$ are the roots of $P(\lambda_m,Q)$.

It is apparent from these brief remarks that the troublesome
term in its current state involves singularities and poles in both
the upper- and lower-half planes.  To proceed, we rewrite the numerator
using the following manipulations:
\begin{eqnarray}
\prod_l(K-i\chi_{1,l})(K&+&i\chi_{2,l})(K+i\chi_{3,l}) \nonumber\\
&=&\left(\frac{1-i\eta vK}{i\eta v}\right)^{N-1} det_{N-1} \left(f(K){\cal I} +
{\cal M}(1) \right)\nonumber \\
&=& \frac{1}{k}\left(\frac{1-i\eta vK}{i\eta v}\right)^{N-1} \left[
det_{N} \left(f(K){\cal I} + {\cal M}(0)\right) 
- det_{N} \left(f(K){\cal I} + {\cal M}(k)\right) \right]
\nonumber \\
&=& \frac{1}{k}\left(\frac{i\eta v}{1-i\eta vK}\right) \left[
\prod_m(K-iq_{1,m})(K+iq_{2,m})(K+iq_{3,m}) \right. \nonumber\\
&\ &\quad\quad\quad\quad\quad\quad\quad\quad\left. - \prod_m(K-iQ_{1,m})(K+iQ_{2,m})(K+iQ_{3,m})\right]
\end{eqnarray}
The first line of this chain employed an identity from \cite{kl_crack1},
Eq.\ (40), relating the numerator to the determinant of a certain 
matrix formed from ${\cal M}_{N-1}(1)$ and the
identity matrix ${\cal I}$ together with the function
\begin{equation}
f(K)=\left[i\eta v K^3 - (1-v^2)K^2\right]/(1-i\eta v K) .
\end{equation}
The second line in this chain, claiming this determinant is equivalent, 
up to a constant factor,  to the difference of two $N\times N$ determinants
can be proven by expanding each of the matrices about the
first row.  The last line reexpresses each of these two determinants
using more identities from \cite{kl_crack1}, Eqs.\ (38-39).  

After these manipulations, our term can be written
\begin{eqnarray}
&ik&\frac{\prod_l(K-i\chi_{1,l})(K+i\chi_{2,l})(K+i\chi_{3,l})}
{\prod_m(K-iq_{1,m})(K+iQ_{2,m})(K+iQ_{3,m})} \nonumber \\
&\ &\quad\quad\quad\quad\quad =
\frac{\eta v}{1-i\eta v K}\left[\prod_m \frac{K-iQ_{1,m}}{K-iq_{1,m}}
- \prod_m \frac{(K+iq_{2,m})(K+iq_{3,m})}{(K+iQ_{2,m})(K+iQ_{3,m})}
\right]
\end{eqnarray}

As the outside factor has a pole at $-i/\eta v$ in the lower-half plane, 
the second term 
only has singularities and poles in the lower-half
plane and so is in the desired form.  The first term is still
mixed and requires further massaging.  The idea is to
subtract out  the unique lower-half plane pole so that what is left
has only upper-half plane poles and zeros. Thus,
\begin{equation}
\frac{\eta v}{1-i\eta v K}\prod_m \frac{K-iQ_{1,m}}{K-iq_{1,m}}=
\frac{\eta v}{1-i\eta v K}\prod_m\frac{\frac{1}{\eta v} + Q_{1,m}}{\frac{1}{\eta v} + q_{1,m}} + g^-
\end{equation}
where now $g^-$ has only upper-half plane poles and zeros.  We will not need
the explicit form of $g^-$ in the calculation.  What we have is now
sufficient to solve for ${\tilde u}^+$, the Fourier transform of the
displacement of the bottom masses in the crack region, $u_1(x)\theta(x)$.  
Using \cite{kl_crack1}, Eq.\ (42), we find
\begin{eqnarray}
0 &=& {\tilde u}^+\prod_m\frac{(K+iq_{2,m})(K+iq_{3,m})}{(K+iQ_{2,m})(K+iQ_{3,m})}
-\frac{i\Delta}{K+i0^+} \prod_m \frac{q_{2,m}q_{3,m}}{Q_{2,m}Q_{3,m}} \nonumber
\\
&\ &\quad + u_1(0) \frac{\eta v}{1-i\eta v K}\left[
\prod_m \frac{1+\eta vQ_{1,m}}{1+\eta vq_{1,m}}
- \prod_m \frac{(K+iq_{2,m})(K+iq_{3,m})}{(K+iQ_{2,m})(K+iQ_{3,m})}\right]
\end{eqnarray}
Solving for ${\tilde u}^+$, we find
\begin{eqnarray}
\label{utilde}
{\tilde u}^+ &=& \frac{i\Delta}{K+i0^+}\prod_m \frac{q_{2,m}q_{3,m}(K+iQ_{2,m})
(K+iQ_{3,m})}{Q_{2,m}Q_{3,m}(K+iq_{2,m})(K+iq_{3,m})} \nonumber\\
&\ &\quad - u_1(0)
\frac{\eta v}{1-i\eta v K}\left[\prod_m \frac{(1+\eta v Q_{1,m})(K+iQ_{2,m})
(K+iQ_{3,m})}{(1+\eta v q_{1,m})(K+iq_{2,m})(K+iq_{3,m})} - 1\right]
\end{eqnarray}
Fourier transforming and evaluating at $x=0^+$ yields
\begin{equation}
u_1(0) = \Delta\prod_m \frac{q_{2,m}q_{3,m}}{Q_{2,m}Q_{3,m}} - 
u_1(0)\left[\prod_m \frac{(1+\eta v Q_{1,m})}{(1+\eta v q_{1,m})} - 1\right]
\end{equation}
so that
\begin{equation}
u_1(0) = \Delta\prod_m \frac{q_{2,m}q_{3,m}(1+\eta v q_{1,m})}
{Q_{2,m}Q_{3,m}(1+\eta v Q_{1,m})}
\end{equation}
Using $u_1(0)=\epsilon$, $\Delta_G=\epsilon\sqrt{kN+1}$ 
and the relations (see \cite{kl_crack1} Eq.\  (43))
\begin{mathletters}
\begin{eqnarray}
\prod_m q_{1,m}q_{2,m}q_{3,m} &=& (\eta v)^{-N} \\
\prod_m Q_{1,m}Q_{2,m}Q_{3,m} &=& (kN+1)(\eta v)^{-N}
\end{eqnarray}
\end{mathletters}
we obtain our desired result
\begin{equation}
\label{cont_n}
\frac{\Delta}{\Delta_G} = \sqrt{kN+1}\prod_m \frac{q_{1,m}(1+\eta v Q_{1,m})}{Q_{1,m}(1+\eta v q_{1,m})}
\end{equation}

The primary benefit of this method of solution over the direct approach
employed in \cite{kl_crack1} is that for the symmetric crack ($k=2$)
we can take the large-$N$ limit.
To do this, we break up the $N-$fold product into two terms. The
first is
\begin{equation}
\Pi_1=\prod_m \frac{(1+\eta v Q_{1,m})}{(1+\eta v q_{1,m})}
\end{equation}
We transform the product into the exponential of a sum over logarithms,
a sum which for $N$ large we can approximate by an integral, via the
Euler-MacLauren Summation Formula (EMSF)\cite{bender}. Thus
\begin{equation}
\ln \Pi_1 \approx \int_0^N dm \left[\ln (1+\eta v Q_{1,m}) - \ln(1+\eta v q_{1,m}) \right]
\end{equation}
Now, for $k=2$, $\Lambda_m=-4\sin^2(\frac{\pi m}{2N+1})$ and
$\lambda_m=-4\sin^2(\frac{\pi (m-1/2)}{2N+1})$. If we define $\alpha=m/N$,
then we see that 
\begin{equation}
q_1(\alpha)=Q_1(\alpha-1/(2N))\approx Q_1(\alpha) - \frac{1}{2N}\frac{dQ_1}{d\alpha}
\end{equation}
The integral is now a total derivative, and so
\begin{equation}
\ln \Pi_1 \approx \frac{1}{2}\int_0^1 d\alpha \frac{d}{d\alpha}\left[\ln (1+\eta v Q_{1,m}) \right] = \frac{1}{2}\left[\ln (1+\eta v Q_{1}(1))-
\ln(1 + \eta v Q_1(0))\right]  
\end{equation}
When $\alpha=1$, $m=N$ and so $\Lambda_m \approx -4$, and so $Q_1(1)$
satisfies
\begin{equation}
0=P(-4,Q_1(1))=\eta v Q_1(1)^3 + (1-v^2)Q_1(1)^2 - 4(1+\eta vQ_1(1))
\end{equation}
Similarly, when $\alpha$ approaches 0, so does $m$ and so also $\Lambda_m$.
This in turn implies that $Q_1(0)=0$. So, finally,
\begin{equation}
\label{pi1_cont}
\Pi_1 \approx [1+\eta v Q_{1}(1)]^{1/2}
\end{equation}

The second factor is slightly more difficult to treat, since the numerator
and denominator both vanish as $m\to 0$.  To handle this, we regularize the
product by multiplying and dividing by $\prod_m \sqrt{\lambda_m/\Lambda_m}$, 
which we can perform analytically.  Then, the regularized product, $\Pi_2^R$,
 can be transformed
to the exponential of an integral of a total derivative, which can be
calculated explicitly.  In detail,
\begin{equation}
\Pi_2^R = \prod_m \frac{Q_{1,m}\sqrt{-\lambda_m}}{q_{1,m}\sqrt{-\Lambda_m}}
\end{equation}
so that
\begin{eqnarray}
\ln \Pi_2^R &\approx& \int_0^1 d\alpha \frac{1}{2}\frac{d}{d\alpha}\ln \frac{Q_1(\alpha)}{\sqrt{-\Lambda(\alpha)}} \nonumber\\
&=& \frac{1}{2}\left[\ln\left(Q_1(1)/2\right) - \ln\left(1/\sqrt{1-v^2}\right)
\right]
\end{eqnarray}
where we have used the fact that for $\alpha$ small, $Q_1(\alpha) \approx
\sqrt{-\Lambda(\alpha)/(1-v^2)}$.
Thus
\begin{equation}
\Pi_2^R \approx \left(\frac{Q_1(1)\sqrt{1-v^2}}{2}\right)^{1/2}   .
\end{equation}
Also,
\begin{eqnarray}
\Pi_2 &=& \Pi_2^R \prod_m \sqrt{\Lambda_m/\lambda_m} \nonumber \\
&=& \Pi_2^R \sqrt{det{\cal M}(2)/det{\cal M}(0)} \nonumber \\
&=& \Pi_2^R \sqrt{2N+1}
\end{eqnarray}
so putting all the pieces together yields the simple result
\begin{equation}
\label{final_cont}
\frac{\Delta}{\Delta_G} = \sqrt{kN+1}\frac{\Pi_1}{\Pi_2 } =
(1-v^2)^{-1/4}\sqrt{\frac{2(1+\eta v Q_1(1))}{Q_1(1)}}
\end{equation}
which expresses $\Delta$ in terms of $Q_1(1)$, the wave-vector at the
end of the Brillioun zone.

The most striking lesson of this formula is that $\Delta$ diverges at
$v=1$, the wave speed.  Thus, while at any finite $N$, there is no
upper limit to the velocity, at infinite $N$ the wave speed is
an absolute upper bound to the crack velocity.  At large $\Delta$, $v$
approaches unity from below as $1/\Delta^4$.  A second lesson is
that at small velocity, 
$\Delta \sim \Delta_G(1 + \eta v)$, so that $\Delta$ approaches
$\Delta_G$ linearly, as is generally true for this $x$--continuum model.
A third implication  is the behavior at large $\eta$.  For fixed
$\Delta$, $v$ decreases as $\eta$ gets large, so that $Q_1(1)$ satisfies
\begin{equation}
0 \approx \eta v Q_1(1)^3 + Q_1(1)^2 - 4(1 + \eta vQ_1(1)) = 
(1+\eta vQ_1(1))(Q_1(1)^2-4)
\end{equation}
so that $Q_1(1)\approx 2$.  Substituting this in Eq.\ (\ref{final_cont}) gives
\begin{equation}
\frac{\Delta}{\Delta_G} \approx \sqrt{1+ 2\eta v}
\end{equation}
or
\begin{equation}
\eta v\approx\left(\frac{\Delta}{\Delta_G}\right)^2 - 1
\end{equation}
In this large $\eta$ limit, of course, $\Delta$ is a function of the
scaling variable $\eta v$, which was first introduced in \cite{sander}.

We have seen how at infinite $N$, the crack speed $v$ never crosses
unity, the wave speed.  However, at any finite $N$, there is
a  $\Delta$ for which the crack speed crosses unity, which
 must diverge with $N$.  We now calculate how this threshold scales
with $N$.  The key to the calculation is $\Pi_2$, since it is the vanishing
of $\Pi_2$ which leads to the divergence of $\Delta$ at $v=1$ for $N$ infinite.
To compute the value of $\Pi_2$ at $v=1$ for finite large $N$, we need to choose a different regularization.  We now define
\begin{equation}
\Pi_2^R= \prod_m \frac{Q_{1,m}(-\lambda_m)^{1/3}}{q_{1,m}(-\Lambda_m)^{1/3}}
\end{equation}
so that
\begin{equation}
\Pi_2 = \Pi_2^R\left[\prod_m \frac{\Lambda_m}{\lambda_m}\right]^{1/3} 
= (2N+1)^{1/3}\Pi_2^R
\end{equation}
Now, since $Q_{1}(\alpha)/(-\Lambda(\alpha))^{1/3}$ approaches the finite limit
$1/\eta^{1/3}$ as $\alpha$ goes to 0,
$\Pi_2^R$ has a finite limit at $v=1$ as $N$ goes to infinity,
namely $\Pi_2^R \approx \sqrt{Q_1(1)(\eta/4)^{1/3}}$.  Using the infinite
$N$ limit of $\Pi_1$ from Eq.\  (\ref{pi1_cont}), we get
\begin{equation}
\frac{\Delta}{\Delta_G} = \sqrt{2N+1}\frac{\Pi_1}{\Pi_2 } \sim
(N)^{1/6}\left[\frac{2(1+\eta Q_{1}(1))}{Q_1(1)}\right]^{1/2}
\end{equation}
Thus, the threshold $\Delta$ scales as $N^{1/6}\Delta_G$, in accord
with the numerical evidence discussed in \cite{kl_crack1}. The coefficient
goes to 2 for $\eta$ large, and vanishes as $\eta^{1/6}$ for small $\eta$.

Another manifestation of this same phenomenon, the disappearance of the
$v=1$ crossing in the infinite-$N$ limit, is the nonuniformity of the
the large-$N$ limit as $v$ approaches $1$.  Working out the corrections
to the EMSF, we find that
\begin{equation}
\Pi_1(N) \approx \Pi_1(N=\infty) * \left(1 - \frac{\pi\eta v}{8\sqrt{1-v^2}N}
\right)
\end{equation}
and
\begin{equation}
\Pi_2^R(N) \approx \Pi_2^R(N=\infty) * \left(1 + \frac{\eta v^3}{16(1-v^2)^{3/2}N}
\right)
\end{equation}
Thus, the relative error of the infinite--$N$ approximation is $O(1/N)$, 
and diverges as $v$ approaches $1$ as $(1-v)^{-3/2}$.  It is also
interesting to note that the relative error vanishes as $v$ goes to zero,
son that the infinite--$N$ approximation becomes better at small velocities.

One last interesting piece of information we can derive from our solution
is the size of the ``process zone'', the region where the solution from
continuum elastic theory breaks down.  The leading-order macroscopic solution
was derived in \cite{kl_crack1}, and exhibited the classic square-root 
singularity at the crack tip, $x=0$.  This singularity in really present only
at infinite $N$, and is cut off by the upper limit on the $Q$'s, (relative
to the smallest $Q \sim 1/N$) at finite $N$.  We can determine the structure 
of the process zone which replaces the singularity by studying our exact
solution for $\tilde u^+$, Eq.\ (\ref{utilde}), for $K$'s of order 1.  Using
the EMSF to evalute the infinite product, similar to the derivations above,
we find
\begin{eqnarray}
{\tilde u}^+ &=& \frac{i\Delta}{\sqrt{2N+1}(K+i0^+)}\left(\frac{Q_1(1)\sqrt{1-v^2}(K+iQ_2(1))(K+iQ_3(1))}{2(K+iQ_2(0))(K+iQ_3(0))}\right)^{1/2} \nonumber \\
&\ & + \frac{i\epsilon}{K+i/\eta v}\left[1-\left(\frac{(1+\eta v Q_1(1))
(K+iQ_2(1))(K+iQ_3(1))}{(1+\eta v Q_1(0))(K+iQ_2(0))(K+iQ_3(0))} \right)^{1/2}\right]
\end{eqnarray}
Using $Q_1(0)=Q_2(0)=0$, $Q_3(0)=(1-v^2)/\eta v$, and the result for $\Delta$,
Eq.\ (\ref{final_cont}), we get
\begin{equation}
{\tilde u}^+ = i\epsilon\left[\frac{(1+\eta v Q_1(1))(K+iQ_2(1))(K+iQ_3(1))}
{(K+i0^+)(K+i(1-v^2)/\eta v)}\right]^{1/2}\left(\frac{1}{K+i0^+}-\frac{1}{K+i/\eta v}\right)
 + \frac{i\epsilon}{K+i/\eta v}
\end{equation}

Examining this expression for small $K$, we find the expected $K^{-3/2}$
singularity, which gives rise to the square-root singularity of the outer
solution.  The coefficient of the $K^{-3/2}$ singularity to leading order 
in $N$
is $\Delta i^{3/2}N^{-1/2}(1-v^2)^{-1/4}$, which reproduces the same $\eta$ independent
coefficient of the square-root singularity, or equivalently stress-intensity
factor, found in \cite{kl_crack1}.
The structure of the process zone is governed by the other singularities in
${\tilde u}^+$ that lie off the origin.  In particular, the
size of the process zone is determined by the
singularity nearest the real line.  For small $\eta$, this is at $K=-iQ_2(1)
\approx -2i/\sqrt{1-v^2}$, so the process zone is truly microscopic, unless the
velocity is very close to 1.  For large $\eta$, the dominant singularity
is at $K=-iQ_3(1)\approx -i/\eta v$, so the process zone grows linearly
with $\eta$ in size.

\section{The Lattice Model}

In this section, we generalize our solution of the continuum model
to the lattice model.  For ease of presentation, we will present the
derivation only in the $N=1$ case.  The case of general $N$ follows
in a straightforward manner from this derivation and that of the continuum
finite $N$ model presented in the previous section.

Our derivation follows directly along the lines of our WH treatment 
of the continuum $N=1$ problem in \cite{kl_crack1}.  
The equation of motion of the steady-state crack is
\begin{equation}
\ddot u(t) = (1 + \eta \frac{d}{dt})\left[u(t + 1/v) - 3u(t)
+ u(t-1/v)\right] - k\theta(-t)(1 + \eta\frac{d}{dt})u(t)
\end{equation}
Upon Fourier transforming, we find
\begin{equation}
0= (1-i\eta v K)\left(4\sinh^2(\frac{iK}{2}) - 1\right){\tilde u}
+ v^2 K^2 {\tilde u} - k(1-i\eta v K){\tilde u}^- + \Delta\delta(K)
- k\eta v u(0)
\end{equation}
where $\tilde u$ is the Fourier transform of $u$ and ${\tilde u}^\pm$
are the transforms of $\theta(\pm t) u(t)$.  We define the function
\begin{equation}
R(\lambda;Q)\equiv(1+\eta v Q) (4\sinh^2(Q/2) + \lambda) - v^2 Q^2
\end{equation}
in terms of which
\begin{equation}
0= R(-(1+k);-iK){\tilde u}^- + R(-1;-iK){\tilde u}^+ 
+ \Delta\delta(K) - k \eta v u(0)
\end{equation}

This function, $R(\lambda;Q)$, 
which is the lattice equivalent of the polynomial $P$ 
employed in the previous section, has not 3 roots, but in fact an infinite set
of zeros in the complex plane.  We shall label these zeros according
to their real parts, $Q_{1,n}$ ($q_{1,n}$) are the zeros of $R(-(1+k);Q)$
($R(-1;Q)$) with positive real parts, and $Q_{2,n'}$ ($q_{2,n'}$) are
their counterparts with negative real parts.  The indices $n$, $n'$ run
over the entire infinite set of zeros but are otherwise left unspecified
for now.  We can decompose $R$ in terms of its zeros
\begin{eqnarray}
R(-(1+k);-iK)&=&-(1+k)\prod_{n,n'} (1 + i\frac{K}{Q_{1,n}})(1 - i\frac{K}{Q_{2,n'}}) \nonumber\\
R(-1;-iK)&=&-\prod_{n,n'} (1 - i\frac{K}{q_{1,n}})(1 + i\frac{K}{q_{2,n'}}) 
\end{eqnarray}

Using this, we rewrite the equation of motion:
\begin{eqnarray}
0&=& -(1+k)\prod_n\frac{q_{1,n}(K-iQ_{1,n})}{Q_{1,n}(K-iq_{1,n})}{\tilde u}^- 
-\prod_{n'}\frac{Q_{2,n'}(K+iq_{2,n'})}{q_{2,n'}(K+iQ_{2,n'})}{\tilde u}^+ 
\nonumber\\
&\ & \quad + \Delta\delta(K) - k \eta v u(0)\prod_{n,n'}\frac{1}{(1-i\frac{K}{q_{1,n}})
(1-i\frac{K}{Q_{2,n'}})}
\end{eqnarray}

As in the last section, the hard part is to decompose the last term.  The
trick is the same, rewriting the numerator as the difference of $R$'s.
\begin{eqnarray}
k\prod_{n,n'}\frac{1}{(1-i\frac{K}{q_{1,n}})(1-i\frac{K}{Q_{2,n'}})} &=&
\frac{1}{1-i\eta v K}\frac{R(-1;-iK)-R(-(1+k);-iK)}{\prod_{n,n'}(1-i\frac{K}{q_{1,n}})(1-i\frac{K}{Q_{2,n'}})} 
\nonumber \\
&=&\frac{1+k}{1-i\eta v K}\prod_n\frac{q_{1,n}(K-iQ_{1,n})}{Q_{1,n}(K-iq_{1,n})}
- \frac{1}{1-i\eta v K}\prod_{n'}\frac{Q_{2,n'}(K+iq_{2,n'})}{q_{2,n'}(K+iQ_{2,n'})}
\end{eqnarray}
As before, the second term is now fine, but the first term is still mixed.
Again we subtract out the unique pole in the lower-half plane which is
what we need to find ${\tilde u}^+$.
\begin{equation}
\frac{1+k}{1-i\eta v K}\prod_{n}\frac{q_{1,n}(K-iQ_{1,n})}{Q_{1,n}(K-iq_{1,n})} 
= \frac{1+k}{1-i\eta v K}\prod_{n'}\frac{q_{1,n}(\frac{1}{\eta v}+Q_{1,n})}{Q_{1,n}(\frac{1}{\eta v}+iq_{1,n})} + g^-
\end{equation}
where $g^-$ only has poles and zeros in the upper-half plane.  Separating
out the pieces analytic in the upper-half plane yields
\begin{eqnarray}
0&=& -\prod_{n'}\frac{Q_{2,n'}(K+iq_{2,n'})}{q_{2,n'}(K+iQ_{2,n'})}{\tilde u}^+
\nonumber\\
&\ & \quad + \frac{i\Delta}{K+i0^+} 
- \eta v u(0)\left[ \frac{1+k}{1-i\eta v K}\prod_{n}\frac{q_{1,n}(\frac{1}{\eta v}+Q_{1,n})}{Q_{1,n}(\frac{1}{\eta v}+iq_{1,n})} - \frac{1}{1-i\eta v K}\prod_{n'}\frac{Q_{2,n'}(K+iq_{2,n'})}{q_{2,n'}(K+iQ_{2,n'})} \right]
\end{eqnarray}
Solving for ${\tilde u}^+$ yields
\begin{eqnarray}
{\tilde u}^+ &=& \frac{i\Delta}{K+i0^+}\prod_{n'} \frac{q_{2,n'}(K+iQ_{2,n'})}
{Q_{2,n'}(K+iq_{2,n'})} \nonumber\\
&\ &\quad - u_1(0)
\frac{\eta v}{1-i\eta v K}\left[(1+k)\prod_{n,n'} \frac{q_{1,n}q_{2,n}(1+\eta v Q_{1,n})(K+iQ_{2,n'})}{Q_{1,n}Q_{2,n}(1+\eta v q_{1,n})(K+iq_{2,n'})} - 1\right]
\end{eqnarray}
Fourier transforming and evaluating at $x=0^+$, we find
\begin{equation}
u_1(0) = \Delta\prod_{n'} \frac{q_{2,n'}}{Q_{2,n'}} -
u_1(0)\left[(1+k)\prod_{n,n'} \frac{q_{1,n}q_{2,n'}(1+\eta v Q_{1,n})}{Q_{1,n}
Q_{2,n'}(1+\eta v q_{1,m})} - 1\right]
\end{equation}
so that
\begin{equation}
\Delta = u_1(0)(1+k)\prod_n \frac{q_{1,n}(1+\eta v Q_{1,n})}
{Q_{1,n}(1+\eta v q_{1,n})}
\end{equation}
As $\Delta_G=u_1(0)\sqrt{1+k}$, we obtain our desired result
\begin{equation}
\frac{\Delta}{\Delta_G} = \sqrt{1+k}\prod_n \frac{q_{1,n}(1+\eta v Q_{1,n})}
{Q_{1,n}(1+\eta v q_{1,n})}
\end{equation}
As there is exactly one real positive root of $R(\lambda;Q)$, it is
convenient to assign this the index 0 and to label the complex roots
in order of imaginary part, so that for example $Q_{1,n}$ and $Q_{1,-n}$ are
complex conjugates.  It is clear the basic structure of the lattice result 
is similar to the continuum result Eq.\ (\ref{cont_n}) above, with
the continuum $Q_{1,1}$, $q_{1,1}$ replaced by their lattice counterparts
$Q_{1,0}$, $q_{1,0}$, and multiplied by a correction factor due to
the additional infinite hierarchy of complex $Q$, $q$'s 
which solve the lattice dispersion relation.

The generalization to finite $N$ is straightforward and is left as an
exercise to the reader.  The result is the direct generalization of the
$N=1$ result.  At finite $N$, there is a set of zeros with positive
real part of $R(\Lambda_m;Q)$, ($R(\lambda_m;Q)$), for each $m=1,\ldots,N$,
now labeled $Q_{1,n,m}$ ($q_{1,n,m}$). Then
\begin{equation}
\frac{\Delta}{\Delta_G} = \sqrt{kN+1}\prod_{n,m} \frac{q_{1,n,m}(1+\eta v Q_{1,n,m})}{Q_{1,n,m}(1+\eta v q_{1,n,m})}
\end{equation}
is the solution to the lattice problem at finite $N$.  It of course
reduces in the limit $\eta \to 0^+$ to the result of Marder and
Gross~\cite{marder_gross}.

As in the continuum, this rather unwieldy formula simplifies tremendously
in the symmetric crack case $k=2$ as $N$ goes to infinity. The procedure
for evaluating the limit is similar to the continuum calculation and so we
do not present the details.  What enters again are $Q_{1,n}(\alpha)$, 
at the two extremes of the Brillouin zone $\alpha=0$, $1$.
If we label $Q_{1,n}(1)=q_{\infty,n}$, $Q_{1,n}(0)=q_{0,n}$ then they
satisfy the dispersion relations
\begin{mathletters}
\begin{eqnarray}
0 &=& R(-4;q_{\infty,n})=(1+\eta v q_{\infty,n}) (4\sinh^2(q_{\infty,n}/2) -4) - v^2 q_{\infty,n}^2 \\
0 &=& R(0;q_{0,n})=(1+\eta v q_{0,n}) (4\sinh^2(q_{0,n}/2)) - v^2 q_{0,n}^2 
\end{eqnarray}
\end{mathletters}

In terms of these $q$'s, the infinite $N$ limit solution is
\begin{equation}
\label{final_lat}
\frac{\Delta}{\Delta_G} = 
(1-v^2)^{-1/4}\sqrt{\frac{2(1+\eta v q_{\infty,0})}{q_{\infty,0}}}\left[\prod_
{n\ne 0}\frac{q_{0,n}(1+\eta v q_{\infty,n})}{q_{\infty,n}(1+\eta v q_{0,n})}\right]^{1/2}
\end{equation}
Again, this is very essentially similar to its continuum counterpart, with
the real lattice wave-vector $q_{\infty,0}$ playing the role of the continuum
wave vector $Q_{1}(1)$, and with a multiplicative correction due to the
presence of complex lattice wave-vectors.  It should also be noted that
this result reduces to that of Slepyan \cite{slepyan} in the $\eta \to 0^+$ limit.

\section{The Small Velocity Limit}
We begin our explorations of the content of our key result, Eq.\ (\ref{final_lat}) by examining the $\eta$ fixed, $v \to 0^+$ limit.  
It is not sufficient to simply set $v=0$, since as $v$ gets smaller, more
and more terms contribute significantly to the infinite product, as seen
in Fig. \ref{v_vs_delta_eta=.5_partial}.  The proper
treatment is to replace the infinite product by an infinite sum of
logarithms and then approximate the infinite sum by an integral via
the EMSF.  Note that for $v=0$, $q_{\infty,n}$ satisfies
$\sinh^2 \frac{q_{\infty,n}}{2} = 1$, with the solution
\begin{mathletters}
\label{q_v0}
\begin{equation}
q_{\infty,n}^{v=0} = 2\pi i n + \omega    ,
\end{equation}
 where $\omega$ is the unique
real root of the equation, namely $\omega=2\ln(1+\sqrt{2})$.
Similarly, 
\begin{equation}
q_{0,n}^{v=0} = 2\pi i n. 
\end{equation}
\end{mathletters}
As we discussed above, we need to consider $v \to 0$, $n \to \infty$,
$\alpha \equiv 2\pi\eta v n$ fixed.  Then, writing 
$q_{\infty,n}\equiv 2\pi i n + \omega_\infty$, $\omega_\infty$ satisfies
\begin{eqnarray}
\sinh^2\frac{\omega_\infty}{2} &=& 1 + \frac{(2\pi i n + \omega_\infty)^2v^2}
{4(1 + \eta v (2\pi i n + \omega_\infty))} \\
&\approx& 1 - \frac{\alpha^2}{4\eta^2(1 + i \alpha)} .
\end{eqnarray}
Similarly,
\begin{equation}
\sinh^2\frac{\omega_0}{2} \approx - \frac{\alpha^2}{4\eta^2(1 + i \alpha)} .
\end{equation}

We can now easily approximate the first infinite product, 
\begin{equation}
\Pi_1 = \prod_{n=-\infty}^\infty \frac{1 + \eta v q_{\infty,n}}{1 + \eta v q_{0,n}} 
\end{equation}
yielding
\begin{eqnarray}
\ln \Pi_1 &\approx&\int_{-\infty}^\infty \frac{d\alpha}{2 \pi \eta v} \left[\ln(1 + i\alpha + \eta v \omega_\infty) - \ln(1 + i\alpha + \eta v \omega_0)\right] \nonumber\\
&\approx&\int_{-\infty}^\infty \frac{d\alpha}{2 \pi} 
\frac{\omega_\infty - \omega_0}{1 + i \alpha} .
\end{eqnarray}

The second product is somewhat trickier, because a naive expansion diverges
at small $\alpha$. We define a regularized product
\begin{equation}
\Pi_2^R \equiv \prod_{n\ne 0} \frac{q_{\infty,n}(2 \pi i n)}{q_{0,n}(2\pi i n + \omega)} ,
\end{equation}
so that (using the product formula for $sinh$, and the fact that
$\sinh(\omega/2)=1$)
\begin{eqnarray}
\Pi_2 &=& \Pi_2^R \prod_{n\ne 0} \frac{2 \pi i n + \omega}{2\pi i n} \nonumber\\
&=& \frac{\sinh(\omega/2)}{\omega/2} \Pi_2^R\nonumber\\
&=& \frac{2}{\omega} \Pi_2^R
\end{eqnarray}
Our regularized product is now easily approximated,
\begin{eqnarray}
\ln \Pi_2^R &\approx& \int_{-\infty}^{\infty} \frac{d\alpha}{2\pi\eta v}\left[\ln\left(\frac{i\alpha + \omega_\infty\eta v}{i\alpha + \omega\eta v}\right) - \ln\left(\frac{i\alpha + \omega_0\eta v}{i \alpha}\right) \right]\nonumber \\
&\approx& \int_{-\infty}^{\infty}\frac{d\alpha}{2\pi}
\frac{\omega_\infty-\omega-\omega_0}{i\alpha}
\end{eqnarray}
Using the identity 
\begin{equation}
\int_{-\infty}^{\infty} d\alpha \frac{\omega}{1+i\alpha} = \pi \omega ,
\end{equation}
we obtain our desired result
\begin{eqnarray}
\label{low_v}
\frac{\Delta|_{0^+}}{\Delta_G} &=& \left[\frac{2\Pi_1}{q_{\infty,0}\Pi_2}\right]^{1/2} \nonumber \\
 &\approx& e^{\omega/4}\exp\left(\frac{i}{2}\int_{-\infty}^{\infty}
\frac{d\alpha}{2\pi} \frac{\omega_\infty-\omega-\omega_0}{\alpha(1+i\alpha)}
\right)
\end{eqnarray}

This result is, as desired, explicitly independent of $v$, but would appear
to depend on $\eta$ through the very nontrivial $\eta$ dependence of 
$\omega_\infty$ and $\omega_0$ under the integral.  It is possible to 
explicitly evaluate the
integral for small and large $\eta$.  For large $\eta$, 
$\omega_\infty-\omega \sim O(1/\eta^2)$ and 
$\omega_0 \sim O(1/\eta)$, so the integral vanishes
and so 
\begin{equation}
\Delta|_{0^+}/\Delta_G = e^{\omega/4} = \sqrt{1 + \sqrt{2}} 
\approx 1.554  .
\end{equation}
For small $\eta$ \cite{slepyan}, $\omega_\infty-\omega_0$ is concentrated at small $\alpha \sim O(\eta)$, so it is appropriate to convert the integral into a
principal value integral and do the $\omega$ integral immediately.
In the remaining integral,
we change variables to $\beta = \alpha/2\eta$.  Then,
the denominator in the integrand reduces to $1/\beta$, so only the
odd (i.e. imaginary) part of $\omega_\infty-\omega_0$ contributes. For
 $\beta \ge 0$ we find 
\begin{equation}
Im\, \omega_\infty = 2 Im\, \sinh^{-1}(\sqrt{1-\beta^2}) = \left\{
\begin{array}{cl}
0 & \beta  \le 1 \\
2\sin^{-1}\sqrt{\beta^2-1}\quad & 1 \le \beta \le 2 \\
\pi & \beta \ge 2 
\end{array}\right.
\end{equation}
and 
\begin{equation}
Im\, \omega_0 = 2 Im\, \sinh^{-1}(\sqrt{-\beta^2}) = \left\{
\begin{array}{cl}
2 \sin^{-1}\beta \quad & \beta \le 1 \\
\pi & \beta \ge 1
\end{array}\right.
\end{equation}
The integral thus becomes
\begin{eqnarray}
\int_{-\infty}^{\infty}
\frac{d\alpha}{2\pi} \frac{\omega_\infty-\omega-\omega_0}{\alpha(1+i\alpha)}
 &=& \frac{i\omega}{2} - 2i\int_0^1 \frac{d\beta}{\pi} 
\frac{\sin^{-1} \beta}{\beta} + 2i\int_1^2 \frac{d\beta}{\pi}
\frac{\sin^{-1} \sqrt{\beta^2-1} - \pi/2}{\beta} \nonumber\\
&=&i \ln(1+\sqrt{2}) - i\ln 2+\frac{i}{2}\ln(\frac{3+\sqrt{8}}{4}) \nonumber\\
&=& 0
\end{eqnarray}
So, again the integral vanishes, and the result for large and small $\eta$
is the same.  One is lead to guess that in fact the integral vanishes
for all $\eta$, as indeed a numerical computation confirms.
This is physically reasonable, since $\Delta|_{0^+}$
should be nothing other than the maximal $\Delta$ for an arrested crack, 
which was previously found numerically \cite{kl_crack1} to be approximately 
$1.55 \Delta_G$. This maximal arrested crack $\Delta$ is the result of a static
calculation, and is of course completely independent of $\eta$.  The
vanishing of the integral can be demonstrated analytically and is the
result of the fact that the integral has no singularities in the lower-half-plane.  One can then close the contour there and the result is identically zero.

To see this, one has to study the analtyic structure of the functions
$\omega_\infty(\alpha)$, $\omega_0(\alpha)$. Consider first 
$\omega_0(\alpha) = 2\sinh^{-1} (y(\alpha))$, where $y^2(\alpha)\equiv
-\alpha^2/(4\eta^2(1+i\alpha))$. Since $sinh^{-1}(y)=2\ln(\sqrt(y^2) + 
\sqrt{1+y^2}$,
$\omega_0$ has a branch cut singularity along the line  $y^2=-r$ where $1>r>0$.
Working out the algebra, in the complex $\alpha$ plane this works out
to be, for $\eta>1$, two separate curves. The first is a segment
along the upper imaginary axis from
$\alpha=2i \eta(\eta - \sqrt{\eta^2-1})$ up to 
$\alpha=2i \eta(\eta + \sqrt{\eta^2-1})$. The second is the
circle of radius 1 centered at the point 
$\eta=i$.  Similarly, $\omega_\infty$ has branch cuts for $2>r>1$, which
are two finite segments along the positive imaginary axis extending above
and below the $\omega_0$ branch cuts.
For $\sqrt{2}/2\eta<1$, the branch cut for $\omega_0$ is a 
sector of the circle,
while the branch cut for $\omega_\infty$ is the rest of the circle
and a finite piece of the
the entire imaginary axis extending centered about $2i$.
For $\eta<\sqrt{2}/2$, the branch cuts are confined entirely to a part of the
circle.
Thus, the singularities for all $\eta$
lie entirely in the upper-half plane
and, as advertised, the integrand is analytic in the lower-half-plane and so
the integral vanishes.

The next step is to extend this calculation to next order in $v$.  There
are two sources for this first correction in $v$.  
One comes from the higher-order velocity dependence of 
$q_{\infty,n}$, $q_{0,n}$.  
The other comes
from the EMSF correction to the replacement of the infinite sum by an integral.
The calculation of the first piece is similar in structure 
to the leading order calculation, just more involved. We expand 
$q_{s,n}\approx 2\pi i n + \omega_s + v\sigma_s$, where the subscript $s=
\infty$, $0$,
and find
\begin{equation}
\sigma_s = \frac{\alpha\omega_s}{2\eta\sinh\omega_s}
\frac{2i-\alpha}{1+i\alpha^2} .
\end{equation}
It is straightforward, though tedious, to substitute this in $\Pi_1$, 
$\Pi_2^R$ and 
expand, giving a multiplicative correction factor of
\begin{equation}
1 + \frac{iv}{2} \int_{-\infty}^\infty \frac{d\alpha}{2 \pi} \left[
\frac{\sigma_\infty - \sigma_0}{\alpha(1+i\alpha)} 
+ i\eta(\omega_\infty^2 - \omega_0^2)\frac{1+2i\alpha}{2\alpha^2(1+i\alpha)^2}
- i\eta\frac{\omega^2}{2\alpha^2}\right]
\end{equation}
The integrand is again a very nontrivial function of $\eta$ and $\alpha$,
but again a miracle occurs and the integral vanishes identically (as
seen by numerical computation) for all $\eta$!  The same analytic
argument as above can be used to prove this point.

\begin{figure}
\centerline{\epsfxsize=3.25in \epsffile{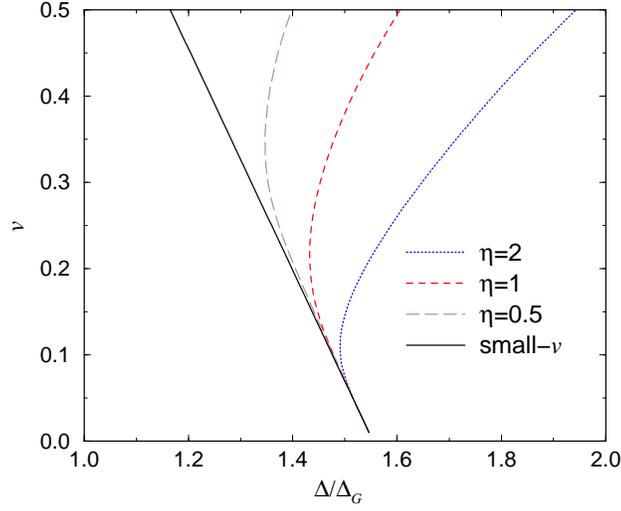}}
\caption{$v$ vs. $\Delta/\Delta_G$ for $\eta=0.5, 1, 2$
along with the asymptotic result for small $v$, Eq.\ (\ref{small_v}).}
\label{small_v_fig}
\end{figure}

This leaves us with only the second source for a $O(v)$ 
correction, namely the first endpoint EMSF correction to the integral.
There are only exponentially small corrections to the integral representation
of $\Pi_1$, but since $\Pi_2^R$ does not include an $n=0$ term, we
need to subtract the $n=0$ limit of the summand, namely 
\begin{equation}
\lim_{n\to 0} \ln(\frac{2\pi i n}{2\pi in+\omega_0})=\ln(\frac{1}{1+v})\approx -v
   ,
\end{equation}
from $\ln \Pi_2^R$. This gives a multiplicative correction factor of $(1+v)$
to $\Pi_2^R$, so we find that for small velocity
\begin{equation}
\label{small_v}
\Delta \sim \Delta|_{0^+} (1-v/2)
\end{equation}
Thus the leading small $v$ behavior of $\Delta$ is {\em completely independent}
of $\eta$.  However, it has $\Delta$ as a strictly decreasing
function of $v$. As we shall see in the next section, the turnaround
for larger $v$ is a nonperturbative effect. For now,
we will conclude this section by showing in Fig. \ref{small_v_fig} 
a plot of the small velocity
region of the graph for various $\eta$'s, together with our analytic
approximation.  We see that the analytic result is
confirmed. 

\section{Large $\eta$ Limit}
We now turn to a study of the large $\eta$ limit.  In this limit,
as first pointed out by Pla, et. al. \cite{sander}, the relevant variable
is $\eta v$.  Thus, we study the limit $\eta \to \infty$, $v \to 0$,
$\phi \equiv \eta v$ fixed. As we shall see, this calculation will
shed much light on the small $v$ results we obtained in the previous
section. 

To begin the calculation, we need the $q_{\infty,n}$'s and
$q_{0,n}$'s at $v=0$ that we obtained in the previous section, Eq.\ (\ref{q_v0}).  Then
\begin{equation}
\Pi_1 = \prod_{n=-\infty}^{\infty} \frac{1+\eta v q_{\infty,n}}{1+\eta v q_{0,n}} = \prod_{-\infty}^{\infty} \frac{1 + \phi (2\pi i n + \omega)}{1 + \phi (2 \pi i n)} = \frac{\sinh(\frac{1+\omega\phi}{2 \phi})}{\sinh(\frac{1}{2\phi})}
\end{equation}
Similarly,
\begin{equation}
\Pi_2 = \prod_{n\ne 0} \frac{q_{\infty,n}}{q_{0,n}} = \prod_{n \ne 0} 
\frac{2\pi i n + \omega}{2\pi i n} = \frac{2}{\omega}\sinh(\frac{\omega}{2})
= \frac{2}{\omega}
\end{equation}
Thus,
\begin{equation}
\label{large_eta}
\frac{\Delta}{\Delta_G} = \left[\frac{2\Pi_1}{q_{\infty,0}\Pi_2}\right]^{1/2} =
\left[\frac{\sinh(\frac{1+\omega\phi}{2 \phi})}{\sinh(\frac{1}{2\phi})}\right]^{1/2} = \left[\coth(\frac{1}{2\phi}) + \sqrt{2}\right]^{1/2}
\end{equation}

We can invert this relation, solving for $\phi$ in terms of $\Delta$,
which yields
\begin{equation}
\phi=\left[\ln\frac{\left(\frac{\Delta}{\Delta_G}\right)^2 - \sqrt{2}+ 1}
{\left(\frac{\Delta}{\Delta_G}\right)^2 - \sqrt{2} - 1}\right]^{-1}
\end{equation}

For large $\Delta$, this approaches $\frac{1}{2}\left[(\Delta/\Delta_G)^2 -
\sqrt{2}\right]$.  This asymptotic result, which is also
presented in Fig. \ref{large_eta_fig}, is to be contrasted with the result of
our continuum calculation, where we found 
$\phi=\frac{1}{2}\left[(\Delta/\Delta_G)^2 - 1\right]$.  Thus, the continuum infinite-$\eta$ calculation for all $\Delta$ 
essentially reproduces the large-$\Delta$ limit of the lattice calculation,
with the correct functional dependence, but with the graph just
shifted down slightly.  It is also worth noting that including just the
$n=0$ term, instead of the whole infinite product, also gives the same result,
with an intercept of $2/\omega$ which is intermediate between the continuum
calculation and the exact asymptotic result.  As $\Delta$ decreases, the
true $\eta=\infty$ curve falls below the asymptotic result, 
so as to intercept the $\Delta$--axis
at $\Delta|_{0^+}$. The approach is singular, as can be seen by
looking at Eq.\ (\ref{large_eta}) for small $\alpha$. We find
\begin{equation}
\Delta \sim \Delta|_{0^+} (1 + \frac{\Delta_G^2}{\Delta|_{0^+}^2}e^{-1/\phi})
\end{equation}
with an essential singularity at small $\phi$.

Examining Fig. \ref{large_eta_fig} more carefully, we see that
our infinite $\eta$ result has failed to capture one of the most
salient features of the finite $\eta$ data, namely the subcritical
nature of the bifurcation from the arrested state.  Instead, it
possesses a (very-) marginally supercritical onset of the moving crack.
To reproduce the subcritical bifurcation from our analytics, 
we need to generate the next order correction in $1/\eta$.

We begin by generating the next order correction to the $q$'s.  We
find that $q_{\infty,n}$ does not change to this order, but now
$q_{0,n} \approx 2\pi i n + \omega_0$ where
\begin{equation}
\omega_0 = \frac{2\pi i n \phi}
{\eta\left(1 + 4\pi^2 n^2\phi^2\right)^{1/4}}e^{-\frac{i}{2}\tan^{-1} 2\pi n \phi}
\end{equation}
This induces a multiplicative correction to $\Delta$ of
\begin{equation}
1 - \sum_{n \ne 0} \frac{\omega_0}{4\pi i n \phi(1+2\pi i n \phi)}
\end{equation}
We are interested in the effect of this correction at small $\phi$, in
which case we are again free to replace the sum by an integral.  If we
add in the $n=0$ term to the sum, the error will be exponentially small
in $1/\phi$.  So, up to exponentially small terms, the
correction for small $\phi$ is (defining $\alpha=2\pi n \phi$)
\begin{equation}
1 - \frac{\phi}{2\eta} + \int_{-\infty}^{\infty} \frac{d\alpha}{2\pi}\frac{1}{\eta(1+\alpha^2)^{1/4}
(1+i\alpha)}e^{-\frac{i}{2}\tan^{-1}\alpha}
\end{equation}
The integral vanishes, as can be seen by a substitution of variables
$x=(1+\alpha^2)^{-1/4}$.  In fact, the integral is nothing
more than the first-order expansion in $1/\eta$ of the integral in Eq.\ 
(\ref{low_v}) which we found vanishes identically in $\eta$.
We are thus left with a correction factor of simply $(1-\phi/2\eta)=(1-v/2)$
up to exponentially small terms.  This is precisely the small $v$ correction
we found in the previous section.  The full behavior to this order for
small $\phi$ is thus
\begin{equation}
\Delta \sim \Delta|_{0^+} (1 + \frac{\Delta_G^2}{\Delta|_{0^+}^2}e^{-1/\phi})(1 - \frac{\phi}{2\eta})
\end{equation}
This has the subcritical bifurcation we are seeking.  As $\phi$ increases
from $0$, $\Delta$ decreases from $\Delta|_{0^+}$ due to the influence
of the second factor, until the exponential kicks in and causes $\Delta$
to turn around and start increasing.  The $\phi$ at which the turn-around
occurs is, for large $\eta$, of order $1/\ln\eta$ (translating to a velocity
of order $1/\eta\ln\eta$) which goes to 0 as $\eta$
goes to $\infty$, but very slowly.  Thus at infinite $\eta$ there is no
turnaround and $\Delta$ strictly increases with $\phi$ as we found in
the zeroth-order calculation at the beginning of this section.  The minimum
$\Delta$ lies, for large $\eta$, an amount of order $1/\eta (\ln\eta)^2$
below $\Delta|_{0^+}$.

Thus we see that it is the subdominant pieces that are responsible
for the increase of $\Delta$ with $v$, while the perturbative pieces
give rise to the subcritical bifurcation.  Analyzing the subdominant
pieces in a little more depth, it is easy to see that for $\eta>\sqrt{2}/2$
the leading subdominant piece goes as $\exp(-1/\eta v)$.  For smaller
$\eta$, the subdominant piece falls less rapidly, and has an oscillating
component, due to the off-axis branch cut assuming dominance.  This picture
is consistent with the numerical evidence.

\section{Stokes Viscosity}
It is worthwhile to contrast the behavior we have seen for Kelvin viscosity
with that which obtains for Stokes viscosity, where the dissipation is
associated with the mass points and not the bonds.  The calculation
in this case is much simpler, since the troublesome $\eta$ term is
not present.  For our purposes, it is sufficient to consider our
$x$-continuum theory, as the conclusions we obtain carry over to the
full lattice model.  The result for ${\tilde u}^+$ is
\begin{equation}
{\tilde u}^+ = \frac{i\Delta}{K+i0^+}\prod_m \frac{q_{2,m}(K+iQ_{2,m})}
{Q_{2,m}(K+iq_{2,m})}
\end{equation}
where now the $Q$'s satisfy the dispersion relation
\begin{equation}
(1-v^2)Q_m^2 - bvQ_m + \Lambda_m=0
\end{equation}
(and the $q$'s the parallel form with $\lambda_m$) and $b$ is the
Stokes viscosity. This
can be seen by a simple limiting procedure applied to Eq.\ (\ref{utilde}),
or by replaying the derivation leading up to Eq.\ (41) of \cite{kl_crack1}
with $b$ instead of $\eta$.  This form of the solution can be
shown to be equivalent to that
obtained by Marder and Gross\cite{marder_gross}.
This result leads to the solution for $\Delta$:
\begin{equation}
\Delta = \epsilon \prod_m \frac{Q_{2,m}}{q_{2,m}}
\end{equation}
We are interested in the large-$N$ limit, which we obtain be
defining the renormalized product
\begin{equation}
\Pi^R = \prod_m \frac{Q_{2,m}(-\lambda_m)}{q_{2,m}(-\Lambda_m)}
\end{equation}
since the $Q_2$'s are linear in $\Lambda$ for small $\Lambda$.  Applying
the EMSF, we find that for large $N$,
\begin{equation}
\Pi^R \approx lim_{\alpha \to 0}\sqrt{\frac{Q_2(1)(-\Lambda(\alpha))}
{Q_2(0)(-\Lambda(\alpha))}} = (b^2v^2 + 16(1-v^2))^{1/4}\sqrt{\frac{bv}{8(1-v^2)}}
\end{equation}
so that
\begin{equation}
\Delta \approx (2N+1)\epsilon (b^2v^2 + 16(1-v^2))^{1/4}\sqrt{\frac{bv}{8(1-v^2)}}
\end{equation}
The key difference between this formula and the parallel one for $\eta$ is
that $\Delta/\epsilon$ is proportional to $N$, and not $N^{1/2}$ as before.
The reason for this is that the Stokes viscosity is most effective at
damping small wavelengths, and so affects the macroscopic stress fields.
The Kelvin viscosity does not damp out small wavelengths and only
acts on short wavelengths.  Another way to see this is to compute
the stress intensity factor, which in the Stokes case is inversely
proportional to $\sqrt{b}$. The driving force required to propagate the
crack is thus much larger in the Stokes case.  In particular, in the
Stokes case there is no macroscopic scaling limit, where things just
scale with the Griffith driving, $\Delta_G$.  For these reasons, we feel
that the Stokes viscosity is not a good model of dissipation for studying
crack propagation.

The only way to obtain a nice macroscopic limit where $\Delta$ scales like
$\Delta_G$ is to artificially scale $b$ with $N$ so that $b=b_0/N$.  However,
this procedure has no physically satisifying motivation, especially when
the Kelvin viscosity model suffers none of these defects.

\section{Concluding Remarks}

We close by making a few comments about this work and prospects for
future extensions.  First it is important to note that the present work
is limited to a consideration of the steady-state crack.  Thus, aside from
general issues of the size of the process zone, the major output of this
problem is the velocity-driving relation.  Here the most striking
qualitative effect of Kelvin
viscosity is near threshold, reducing the extent of the backward
bifurcation.  Significantly above threshold,
the major role of viscosity is to provide a velocity scale, 
so that the crack velocity becomes inversely proportional to the viscosity.
It is important to understand how viscosity impacts on the stability of the
crack.  It is clear, as Marder and Gross have pointed out\cite{marder_gross}, that
the steady-state crack is unstable in the regime of the backward bifurcation.
The more interesting question is in the higher-velocity regime.  Here,
no systematic studies have been done to examinne the role of viscosity.
It is not clear that the piecewise-linear model considered here is altogether
appropriate for studies of stability, as instabilities can be masked by
inconsistencies of the steady-state solution.  We look forward to reporting
on work in this direction soon, along with generalization to the problem
of Mode I cracking.

\acknowledgments
The author acknowledges useful conversations with H. Levine and
acknowledges the support of the Israel Science Foundation and the
hospitality of the Prof. A. Chorin and the Lawrence Berkeley National Laboratory. The work
was also supported in part by the Office of Energy Research,
Ofice of Computational  and Technology Research, Mathematical, Information 
and Computational Sciences Division, Applied Mathematical Sciences Subprogram,
of the U.S. Department of Energy, under Contract No. DE-AC03-76SF00098.

\end{document}